\documentclass[10pt,conference]{IEEEtran}
\usepackage{amsmath,amsfonts}
\usepackage{algorithmic}
\usepackage{algorithm}
\usepackage{array}
\usepackage[caption=false,font=normalsize,labelfont=sf,textfont=sf]{subfig}
\usepackage{textcomp}
\usepackage{stfloats}
\usepackage{url}
\usepackage{verbatim}
\usepackage{graphicx}
\usepackage{multirow}
\usepackage[usenames,dvipsnames]{color}
\usepackage{booktabs}
\usepackage{cite}
\usepackage[acronym]{glossaries}

\newacronym{ofdm}{OFDM}{orthogonal frequency-division multiplexing}
\newacronym{ls}{LS}{least-squares}
\newacronym{tsne}{$t$-SNE}{$t$-distributed stochastic neighbor embedding}
\newacronym{music}{MUSIC}{multiple signal classification}
\newacronym{dbscan}{DBSCAN}{density-based spatial
clustering of applications with noise}
\newacronym{sumo}{SUMO}{simulation of urban mobility}
\newacronym{csi}{CSI}{channel state information}
\newacronym{dnt}{DNT}{digital network twin}
\newacronym{ve}{VE}{vehicular equipment}

\title{Chartwin: a Case Study on Channel Charting-aided \\ Localization in Dynamic Digital Network Twins}

\author{
\IEEEauthorblockN{Lorenzo Cazzella\IEEEauthorrefmark{1}, Francesco Linsalata\IEEEauthorrefmark{1}, Mahdi~Maleki\IEEEauthorrefmark{1}, \\Damiano Badini\IEEEauthorrefmark{2}, Matteo Matteucci\IEEEauthorrefmark{1}, and Umberto Spagnolini\IEEEauthorrefmark{1}}

\IEEEauthorblockA{\IEEEauthorrefmark{1}Politecnico di Milano, Milan, Italy} 
\IEEEauthorblockA{\IEEEauthorrefmark{2}Huawei Technologies Italia S.r.l., Milan, Italy}

E-mails: \{lorenzo.cazzella, francesco.linsalata, mahdi.maleki, matteo.matteucci, umberto.spagnolini\}@polimi.it \\ \{damiano.badini\}@huawei.com

}

\begin{document}
\maketitle

\begin{abstract}
Wireless communication systems can significantly benefit from the availability of spatially consistent representations of the wireless channel to efficiently perform a wide range of communication tasks. Towards this purpose, channel charting has been introduced as an effective unsupervised learning technique to achieve both locally and globally consistent radio maps. In this letter, we propose \textit{Chartwin}, a case study on the integration of localization-oriented channel charting with dynamic Digital Network Twins (DNTs).
Numerical results showcase the significant performance of semi-supervised channel charting in constructing a spatially consistent chart of the considered extended urban environment. The considered method results in $\approx 4.5$ m localization error for the static DNT and $\approx 6$~m in the dynamic DNT, fostering DNT-aided channel charting and localization.
\end{abstract}


\IEEEpeerreviewmaketitle

\section{Introduction}\label{sec:introduction}

\IEEEPARstart{T}HE construction of spatially consistent representations of the wireless channel can considerably improve the performance and efficiency of next-generation wireless communication systems. Towards this purpose, channel charting \cite{studer2018channel, LatentWireless} has been introduced as an unsupervised learning technique that aims to build a radio map of the wireless communication environment starting from a collection of \gls{csi} data at the BS's. The target radio map is required to have a low dimensionality with respect to the input CSI features space and to consistently represent the spatial properties of the environment in which the communication users move both locally and globally. The achieved low-dimensional radio map can be exploited to effectively solve various communication-centric tasks, like user localization, network planning, blockage hand-over, and network scheduling. The construction of a radio map using channel charting can benefit from different unsupervised learning techniques, among which are auto-encoding deep neural networks, dimensionality reduction and manifold learning. 

Among the works in the literature developed from \cite{studer2018channel}, several branch out towards the development of localization techniques. Among them, \cite{deng2021semi} investigates semi-supervised localization based on the construction of a channel chart consistent with the spatial features of the scenario in which the users are located \cite{deng2018multipoint}. As detailed in Section \ref{sec:method}, this procedure requires that for part of the training dataset, the labelling of the CSI features with the corresponding users' positions is available. When used during training, the labelled features allow the \textit{alignment} of the learned representations with the user positions in the considered scenario.

\Glspl{dnt} are faithful multi-tier representations of wireless communication networks \cite{AlkhateebLoc, CazzellaVTC}. Thanks to their accurate representation of the environment dynamics and detailed modelling of physical interactions with the propagation medium, DNTs can be used to extract \textit{a priori} channel features such as angles of arrival and departure (DoA/DoD), path delays, and geometric Doppler shifts for each point in the environment. These features can then serve as \textit{labelled data} for supervised or semi-supervised training in downstream tasks like channel charting and localization.

In this work, we present the following main contributions:
\begin{itemize} 
    \item Introduce \textit{\textit{Chartwin}}, a case study on the construction of localization-oriented channel charts leveraging the channel spatial and temporal structure from dynamic DNTs.
    
    \item Develop a high-fidelity DNT of a large urban area in Milan, Italy, accurately modelling both the static infrastructure (buildings and walls) and dynamic vehicular meshes in diverse LoS/NLoS conditions using realistic traffic and ray-tracing simulations.
    
    \item Provide a comprehensive numerical evaluation of \textit{Chartwin}, analyzing localization and charting performance under both static and dynamic DNT settings, across position supervision levels ranging from 5\% to 50\% of the training dataset. Results are compared against fingerprinting-based methods, showing substantial gains in spatial consistency and accuracy. 
\end{itemize}
To the best of our knowledge, \textit{Chartwin} is the first study to utilize DNT-based channel charting to enable localization in large and geometrically complex urban environments including accurate vehicular meshes modelling.

\section{System Model}\label{sec:system_model}

We consider an uplink V2I communication system over a bandwidth $B$ where the transmitter and receiver are equipped with rectangular antenna arrays with $N_{T}$ and $N_{R}$ antenna elements, respectively. After synchronization in time and frequency and after the removal of the cyclic prefix, the received signal is given by the convolution between the MIMO channel matrix $\mathbf{h}(t) \in \mathbb{C}^{N_{R}\times N_{T}}$ and the Tx signal $\mathbf{x}(t) \in \mathbb{C}^{N_{T}\times 1}$.

Assuming a sampling step $T = 1/B$ with $t = wT$, the resulting discretized received signal is
\begin{equation}\label{eq:discrete_rx_signal}
    \mathbf{y}[w] = \mathbf{h}[w] * \mathbf{x}[w] + \mathbf{n}[w],
\end{equation}
for $w \in \{1, \dots, W\}$, where $\mathbf{n}(w) \in \mathbb{C}^{N_{R}\times 1}$ is an additive signal distortion term accounting for interference and thermal noise, and $W = \left\lceil \tau_{\text{max}} / T \right\rceil$ is the maximum number of temporal channel taps, determined by the maximum channel temporal support $\tau_{\text{max}}$. For the considered channel estimation task, $\mathbf{x}[w]$ is a known pilot sequence randomly generated and uncorrelated in time and space, i.e., $\mathbb{E}\left[\mathbf{x}[w] \mathbf{x}[\ell]^{\mathrm{H}}\right] = \sigma^2_x \mathbf{I}_{N_T} \delta_{w-\ell}$ for any $w, \ell \in \{1, \dots, W\}$, where $\sigma^2_x$ is the Tx signal power. We model the distortion term $\mathbf{n}[w]$ as a zero-mean random vector with circular Gaussian distribution uncorrelated in time and frequency and generally correlated in space ($\mathbf{Q}_n=\mathbb{E}\left[\mathbf{n}[w] \mathbf{n}[w]^{\mathrm{H}}\right]$ is generally non-diagonal) owing to possible directional interference.

\subsection{Channel model}\label{sec:channel_model}
The discrete-time mmWave channel response is modelled as a sum over $P$ clustered propagation paths:
\begin{equation}\label{eq:channel_matrix_time}
   \hspace{-0.2cm}\mathbf{h}[w] = \sum_{p=1}^{P}\beta_p \, e^{j 2 \pi \nu_p t}\,\mathbf{a}_R(\boldsymbol{\vartheta}_p)\mathbf{a}_T^{\mathrm{T}}(\boldsymbol{\psi}_p)g\left[(w-1)T-\tau_p\right],
\end{equation}
where the $p$th path complex gain $\beta_p$ depends on the types of interaction with the propagation environment (e.g., reflection, diffraction, diffuse scattering) and the path-loss; the path amplitudes $\alpha_p = \beta_p \, e^{j 2 \pi \nu_p t}$ are assumed to follow the Wide-Sense Stationary Uncorrelated Scattering (WSSUS) model, i.e., $\alpha_p \sim \mathcal{CN}\left(0,\Omega_p\right)$. The vectors $\mathbf{a}_T(\boldsymbol{\psi}_p)\in\mathbb{C}^{N_T \times 1}$ and $\mathbf{a}_R(\boldsymbol{\vartheta}_p)\in\mathbb{C}^{N_R \times 1}$ denote the transmit and receive array responses associated with the direction of departure (DoD) $\boldsymbol{\psi}_p$ and direction of arrival (DoA) $\boldsymbol{\vartheta}_p$, respectively. The function $g\left[(w-1)T-\tau_p\right]$ represents the sampled pulse-shaping waveform evaluated at time $t = (w-1)T$ and delayed by the path delay $\tau_p$.
In the context of 5G cellular networks, positioning has advanced significantly with the introduction of 3GPP Release 16, which laid the foundation for commercial location-based services. To this end, dedicated positioning pilot signals have been introduced, namely Position Reference Signals (PRS) in the downlink and Sounding Reference Signals (SRS) in the uplink. Periodically, the BS retrieves the channel over time and frequency by receiving \gls{ofdm}-based pilot symbols from the Vehicular Equipments (VEs), denoted as $\hat{\mathbf{h}}_{t,k}[w, f_n]$, where $w \in {1, \dots, W}$ and $f_n \in {f_1, \dots, f_{N_c}}$, with $t$ representing the current time index and $k$ identifying the VE.

\section{Dynamic digital network twins}\label{sec:dynamic_dt_construction}

\begin{figure}[t!]
    \centering
    \includegraphics[width=.433\textwidth]{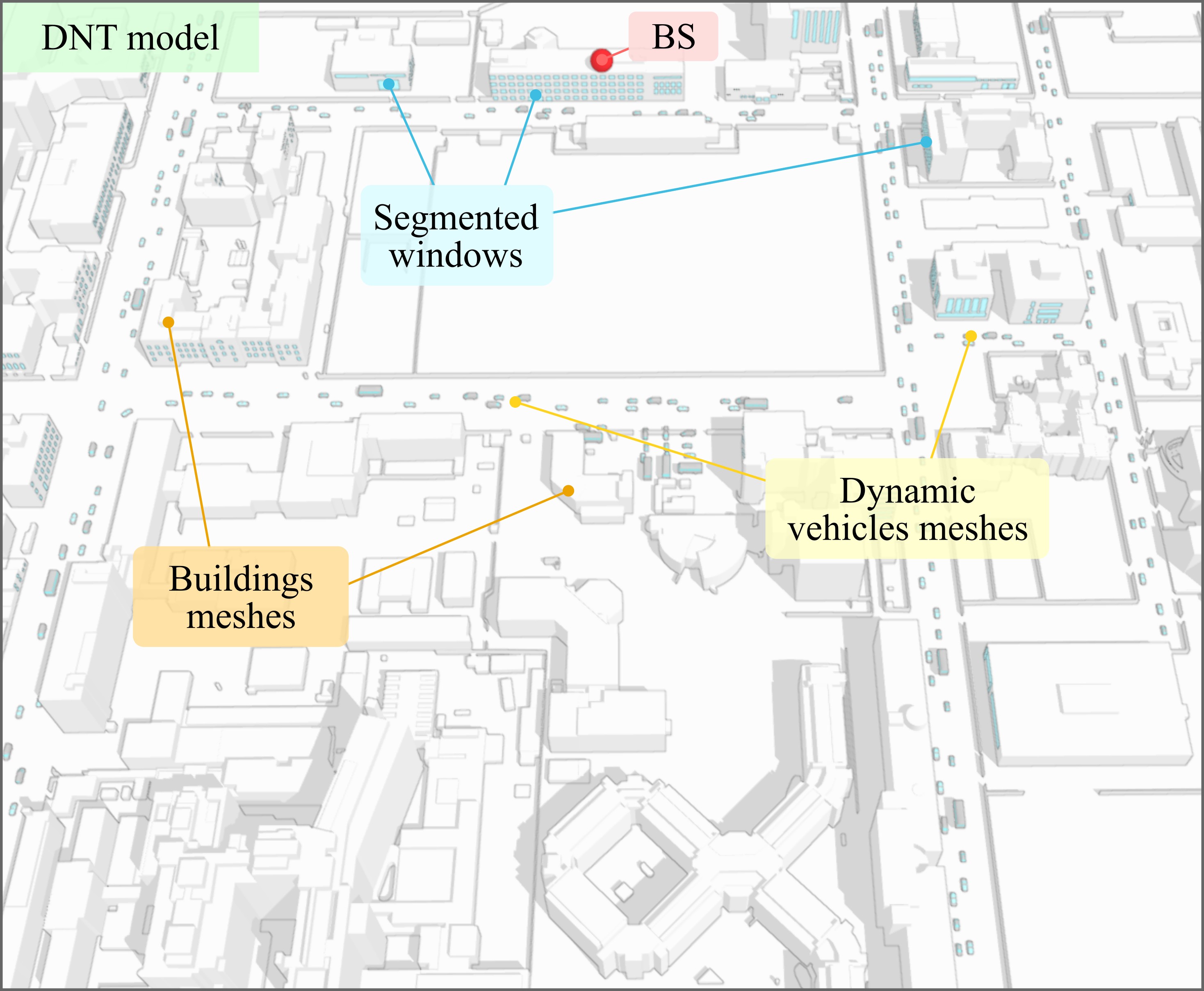}
    \caption{Front view of the modelled DNT of an urban area of Milan, Italy (the area is $\approx 550 \times 670$ m). The red point indicates the position of the BS, while the VEs are located above the 3D vehicles' meshes depicted across the roads.}
    \label{fig:dnt_3d_model}
\end{figure}

In this section, we detail the construction of the dynamic DNT adopted in this work. We consider the high-fidelity 2.5D model of an urban area of Milan, Italy with an extension of $\approx 550 \times 670$ m depicted in Fig. \ref{fig:dnt_3d_model}. The model contains accurate 3D representations of buildings and walls in the real-world environment. To improve the accuracy of the model, we segmented the buildings meshes separating the windows from the rest of the buildings. This allowed us to assign two different radio materials for windows and building structures, i.e., ITU glass and concrete from the ITU recommendation \cite{ITU-R-2040}, respectively.

\textbf{Buildings windows segmentation}. The segmentation of the buildings meshes to separate the windows faces from the rest of the building on the building facades has been articulated in the following steps: we collected 616 high-definition images of the buildings' facades of all the buildings visible from the streets in the considered scenario. After preprocessing the images to remove irrelevant interference factors, we corrected the perspective of the building's facades and stitched multiple camera images to reconstruct large facades captured through multiple acquisitions. We applied the corrected images as textures to the corresponding facades' faces and, finally, we manually segmented the windows mesh components using the Blender 3D modelling software tools.

\textbf{Vehicular traffic simulation}. To enable vehicular traffic simulation over the modelled scenario, we gathered the traffic network topology information for the area from OpenStreetMap (OSM) and we imported it within the \gls{sumo} \cite{SUMO2018} vehicular traffic simulator. We set the SUMO simulation parameters to achieve a realistic vehicle flow in the urban environment. For each simulation time step, we retrieved from SUMO the vehicle classes, vehicle IDs, positions and orientations of the simulated vehicles.

\textbf{Dynamic Ray tracing channel simulation}. We selected four different vehicle meshes from the CARLA \cite{Carla} automotive simulator: sedan, hatchback, truck, and bus. We segmented the vehicles' meshes into vehicle bodies, lights, windows, and rims to assign different radio materials to the separated components. The perfect electric conductor (PEC) radio material has been assigned to the vehicles' bodies and rims, while glass has been set for windows and lights. The vehicle mesh types have been used to model the dynamic part of the scenario. We used the vehicular traffic information retrieved from SUMO to update the radio propagation environment in NVIDIA Sionna by moving the vehicles' meshes according to the gathered positions and orientations and setting the corresponding VEs. The VEs are positioned $0.2$ m above the vehicles' rooftops at different heights depending on the vehicle ($1.67$ m for sedans, $1.8$ m for hatchbacks, $2.76$ m for trucks, and $4.41$ m for buses).

For different VE positions and its associated multipath component $p$, the dynamic DNT provides the tuple $\left(\boldsymbol{\psi}_p, \boldsymbol{\vartheta}_p, \tau_p, \nu_p, |\beta_p|^2 \right)$. 
These parameters define the geometric and temporal structure of the channel in \eqref{eq:channel_matrix_time} and can be considered as a priori information available from the DNT for any given transmitter-receiver pair and spatial location. Such prior knowledge enables geometry-informed modelling and learning strategies by constraining the space of possible channel realizations to those consistent with the environmental layout and mobility conditions encoded in the DNT.

\begin{figure}[t!]
    \centering
    \includegraphics[width=.433\textwidth]{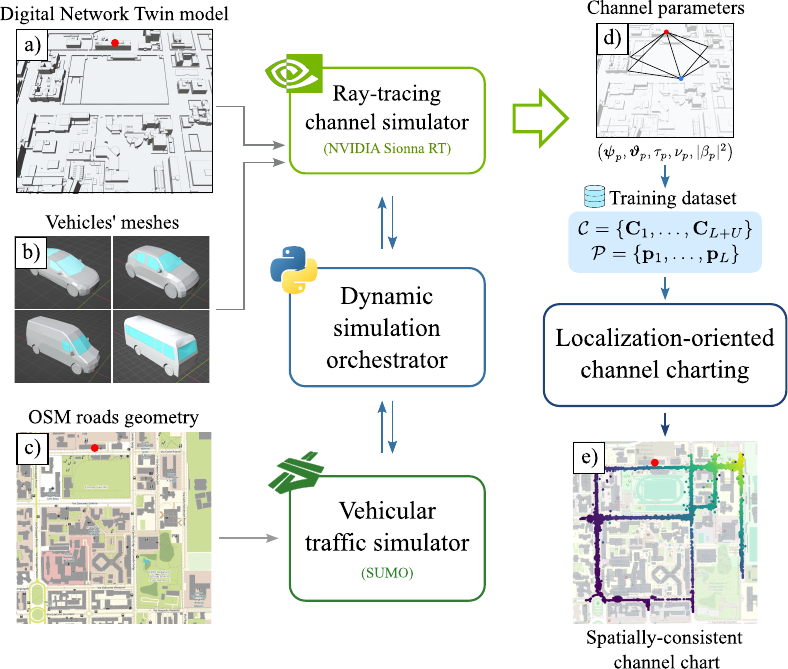}
    \caption{Proposed \textit{Chartwin} case study, integrating the DNT construction workflow with localization-oriented channel charting to build spatially consistent charts of the radio propagation environment. The buildings and vehicle meshes are processed and imported into NVIDIA Sionna. The simulation is updated according to the vehicular traffic data from SUMO.}
    \label{fig:dnt_construction_and_model}
\end{figure}

We provide in Fig. \ref{fig:dnt_construction_and_model} a graphical representation of the DNT construction workflow, wherein a) and b) are the DNT and the vehicles' meshes, respectively, used as inputs to the ray-tracing channel simulator; c) depicts the scenario with the roads information extracted from OSM and provided as input to SUMO; d) represents the channel parameters produced by the ray tracer, while e) shows the positions of a set of test VEs (not used at training time), where the diagonal colour gradient represents the consistency between the position of the chart point with respect to the VE ground truth locations.

\section{Semi-supervised channel-based localization}\label{sec:method}

In this section, we detail the \gls{tsne}-based semi-supervised localization model proposed in \cite{deng2021semi}, which we adopt as a joint channel charting and localization method to assess \textit{Chartwin} on the developed DNT model. Semi-supervision combines the advantages of unsupervised learning, routinely used for channel charting, with limited position supervision to guide the construction of a spatially consistent chart.

We consider a set of VEs moving within an urban environment in the coverage area of a BS. The number and the identities of the VEs vary as they cross the scenario and, therefore, depend on the simulation time step $t$.

\renewcommand{\arraystretch}{1.5}
\begin{table}[t!]
\centering
\footnotesize
\caption{Simulation parameters}
\begin{tabular}{l c | l c }
\toprule
\textbf{Parameter} & \textbf{Value} & \textbf{Parameter} & \textbf{Value}\\
\hline
Carrier freq. ($f_0$) & $28$ GHz & BS ant num. ($N_R$) & $32$ ($4 \times 8$) \\
Bandwidth (B) & $200$ MHz & VE and num. ($N_T$) & $8$ ($4 \times 2$)\\
BS height ($h_{\text{BS}}) $ & $21.7$ m & Points num. ($N$) & $10^4$\\
\bottomrule
\end{tabular}
\label{tab:simulation_parameters}
\end{table}
\renewcommand{\arraystretch}{1}

The CSI sample used for localization and charting is the frequency-domain covariance $\mathbf{C}_k = \mathbb{E}_f[\hat{\mathbf{h}}_{t, k} \hat{\mathbf{h}}_{t, k}^H]$, which encodes angular, temporal, and power information that varies slowly with the position of the VE. We assume that $N$ CSI samples are collected at a single BS, and are divided into a labelled set $\mathcal{L}$ and an unlabeled set $\mathcal{U}$:
\begin{equation}
    \mathcal{L} = \{\mathbf{C}_1, \dots, \mathbf{C}_L\},\quad \mathcal{U} = \{\mathbf{C}_{L + 1}, \dots, \mathbf{C}_{L + U}\}.
\end{equation}
The labeled samples $\mathcal{L}$ and the associated 2D position coordinates $P_{\mathcal{L}} = \{\mathbf{p}_1, \dots, \mathbf{p}_L\}$ are generated from the DNT as described in Section~\ref{sec:dynamic_dt_construction}. For each labelled sample, the tuple $\left(\boldsymbol{\psi}_p, \boldsymbol{\vartheta}_p, \tau_p, \nu_p, |\beta_p|^2\right)$ is available, providing a geometry-consistent ground truth representation of the channel. These are used to construct the CSI features employed in the semi-supervised learning process. Conversely, the unlabeled samples $\mathcal{U}$ are derived from channel estimates $\hat{\mathbf{h}}$ obtained via pilot-based measurements and processed into corresponding covariance matrices. Their associated VE positions $\hat{\mathbf{p}}_{L+1}, \dots, \hat{\mathbf{p}}_{L+U}$ are unknown and must be inferred via the learned chart. The total number of samples is $N = L + U$.

The aim of \gls{tsne} is to embed the high-dimensional features $\mathbf{C}_k$ into low-dimensional data points $\mathbf{z}_k$ such that similar data points are mapped close together, and dissimilar ones are mapped far apart, according to a selected pairwise similarity. The similarity between the input samples is encoded in a symmetric probability matrix $(\mathbf{P})_{i, j} = \frac{1}{2} (p_{i|j} + p_{j|i})$, $\mathbf{P} \in \mathbb{R}^{N\times N}$, where $p_{i|j}$ is the conditional probability that the $i$th sample is a neighbor of the $j$th, modeled via Gaussian kernels. The similarities in the low-dimensional representation are gathered in matrix $\mathbf{Q} \in \mathbb{R}^{N\times N}$, whose elements follow a Student-$t$ distribution. The embedding is obtained by minimizing the Kullback-Leibler (KL) divergence between $\mathbf{P}$ and $\mathbf{Q}$.
\begin{equation}\label{eq:kl_div_objective}
    \mathrm{KL}(P \| Q) = \sum_{i=1}^{N} \sum_{j=1}^N p_{i,j} \log \frac{p_{i,j}}{q_{i,j}}.
\end{equation}
To incorporate supervision, a constraint has been introduced to enforce the mapping of labeled CSI features to the known positions $P_{\mathcal{L}}$ obtained from the DNT by minimizing the KL-divergence in \eqref{eq:kl_div_objective} during training. We refer the reader to \cite{deng2021semi} for further details on \gls{tsne}-based semi-supervised localization.

To compare the CSI features in the high-dimensional space, we computed the dissimilarity measure $d(\mathbf{C}_m, \mathbf{C}_n)$ between features by adopting the Log-Euclidean distance:
\begin{equation}
    d(\mathbf{C}_m, \mathbf{C}_n) = \|\log \mathbf{C}_m - \log \mathbf{C}_n \|_F,
\end{equation}
which has been shown to better preserve global geometry and to produce improved channel charting performance in \cite{kazemi2023beam}.

The complexity of semi-supervised \gls{tsne} \cite{deng2021semi} is $\mathcal{O}(N^2)$ per iteration. The convergence rate depends on the perplexity, momentum, learning rate and on the number of iterations hyperparameters, which we reported in Section \ref{sec:results_charting_localization}.

\section{Numerical results}\label{sec:results}

In this section, we provide the simulation results achieved by applying semi-supervised \gls{tsne} channel charting and localization on the developed dynamic DNT environment. First, we briefly discuss the adopted evaluation metrics. We then evaluate the considered localization method in both static and dynamic environmental conditions. 
The simulation framework is described in Section \ref{sec:dynamic_dt_construction}, which details the construction of the static and dynamic DNTs used in these simulation results. 
We report in Table \ref{tab:simulation_parameters} the system and simulation parameters, assuming, as an example, the usage of mmWave 5G.

\begin{table}[t!]
\centering
\footnotesize
\caption{Localization and channel charting performance varying the supervision level for the static and dynamic DNTs.}
\begin{tabular}{l c c c c c}
\toprule
\textbf{Scenario} & \textbf{Sup.} (\%) & \textbf{CT} $\uparrow$ & \textbf{KS} $\downarrow$ & \textbf{TW} $\uparrow$ & \textbf{Loc. error}\\
\noalign{\smallskip}
\hline
\noalign{\smallskip}
\multirow{5}{*}{Static DNT} & 5 \% & $0.92$ & $0.22$ & $0.92$ & $50.88$ m\\\noalign{\smallskip}
 & 10 \% & $0.96$ & $0.15$ & $0.96$ & $31.39$ m\\\noalign{\smallskip}
 & 25 \% & $0.98$ & $0.09$ & $0.98$ & $12.52$ m\\\noalign{\smallskip}
 & 35 \% & $0.99$ & $0.07$ & $0.99$ & $8.21$ m\\\noalign{\smallskip}
 & 50 \% & $1.00$ & $0.04$ & $1.00$ & $4.50$ m\\
\midrule
\multirow{5}{*}{Dynamic DNT} & 5 \% & $0.91$ & $0.24$ & $0.91$ & $57.18$ m\\\noalign{\smallskip}
 & 10 \% & $0.94$ & $0.18$ & $0.94$ & $36.54$ m\\\noalign{\smallskip}
 & 25 \% & $0.97$ & $0.12$ & $0.97$ & $17.36$ m\\\noalign{\smallskip}
 & 35 \% & $0.98$ & $0.10$ & $0.98$ & $12.28$ m\\\noalign{\smallskip}
 & 50 \% & $0.99$ & $0.06$ & $0.99$ & $5.99$ m\\

\bottomrule
\end{tabular}
\label{tab:loc_and_charting_performance}
\end{table}

\subsection{Evaluation metrics}

\textbf{Channel charting}. We consider the continuity, trustworthiness and Kruskal stress as metrics to quantify the consistency of the channel features within the learned chart with respect to the spatial features of the ambient space. Continuity measures the consistency between the neighborhoods in the ambient space and the ones mapped into the chart. Conversely, trustworthiness measures the presence of new neighbors in the representation space that are not consistent with the ones in the ambient space. Both continuity and trustworthiness assess the \textit{local} consistency of the chart. Kruskal stress (KS) evaluates, instead, differences in the global geometry of the learned chart with respect to the ambient space, i.e., whether distances are preserved in the learned representation. We refer to \cite{hanan2022channel} for a detailed definition of the considered metrics.

\textbf{Localization error}. The considered method employs semi-supervision to learn a chart representation consistent with the VEs positions in the ambient space. Therefore, we evaluate the positioning error through the Euclidean distance between the ground truth VE position and the associated chart point.

\subsection{Channel charting and localization results}\label{sec:results_charting_localization}

\begin{figure}[!t]
    \centering
    \includegraphics[width=\columnwidth]{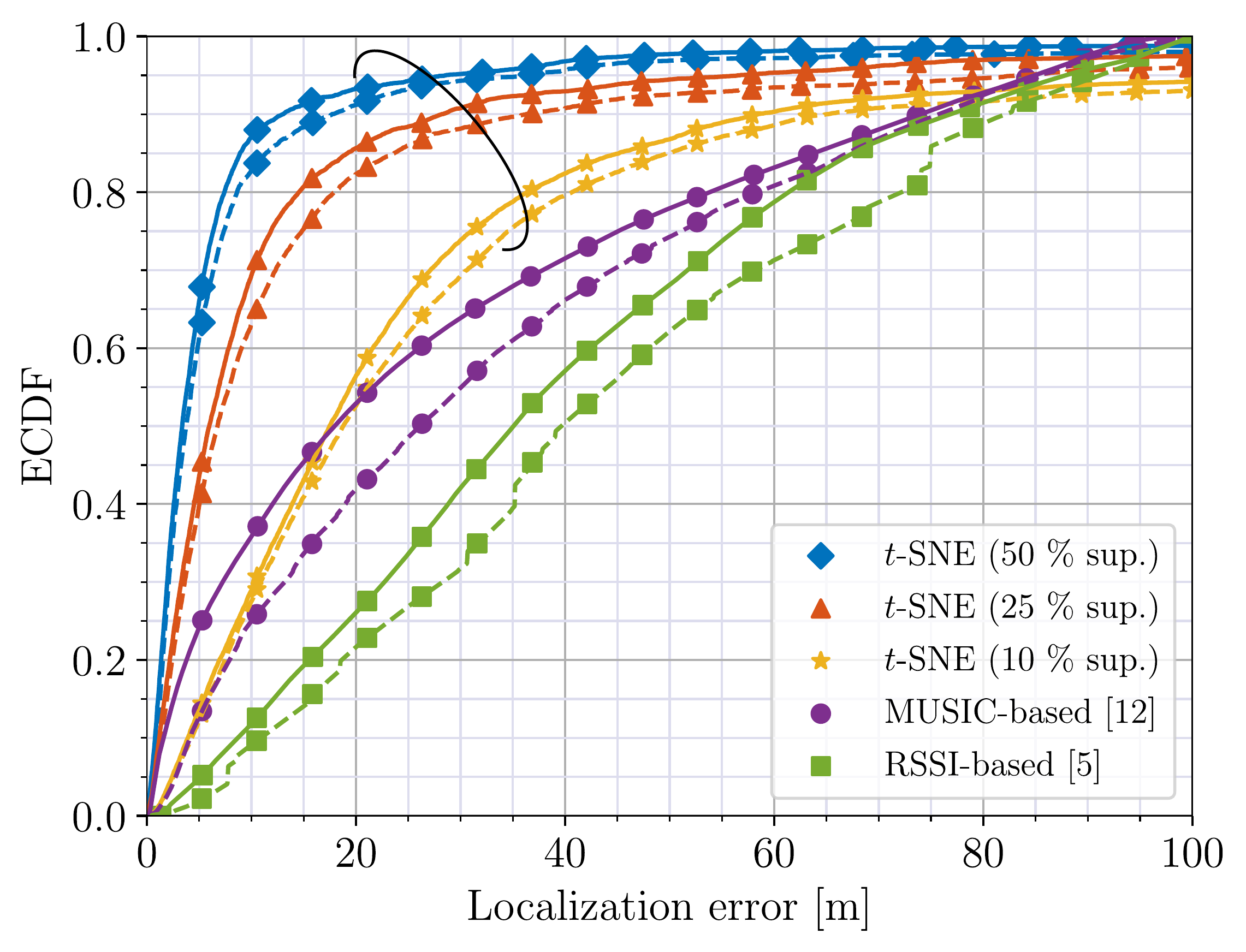}
    \caption{ECDF of the localization error for static (solid lines) and dynamic (dashed lines) DNT conditions. The curves show the localization performance on the validation set for different percentages used to train the selected $t$-SNE based semi-supervised localization method \cite{deng2021semi} and for the comparison methods based on RSSI \cite{AlkhateebLoc} and MUSIC \cite{mazokha2023mobloc}.}
    \label{fig:localization_error_ecdf}
\end{figure}

We propose here the numerical results for the channel charting and localization performance of the selected semi-supervised method discussed in Section \ref{sec:method} in terms of continuity (CT), Kruskal stress (KS) and trustworthiness (TW). The \gls{tsne} model performance is evaluated on a realistic DNT of an urban environment constructed according to the workflow detailed in Section \ref{sec:dynamic_dt_construction}. We have randomly split the simulated \gls{ve} trajectories using part of them---which we refer to as supervision percentage---to train the \gls{tsne} model and 50 \% of them for validation. The total number of data points is $N = 10^4$, uniformly sampled from a dataset of $\approx 1.7 \cdot 10^5$ data points, of which $\approx 40$ \% are in LoS and the remaining are in NLoS condition. The proposed channel charting and localization performance have been evaluated on the validation set over both LoS and NLoS conditions. 
For semi-supervised \gls{tsne}, we determined the perplexity $k_t=400$, momentum $\alpha=0.6$, learning rate $\eta = 100$ and the number of iterations $T = 1500$ hyperparameters as suitable values yielding significant localization performance.

Table \ref{tab:loc_and_charting_performance} summarizes the charting and localization performance in terms of the average CT, KS, TW and localization error over 2D coordinates---i.e., without considering the height of the vehicles in the estimation. We compared performance over the static and dynamic DNTs considering a set of supervision levels in terms of percentages of the dataset (between 5 \% and 50 \%) for which ground truth position labels are used. In the static DNT, there are no moving vehicles' meshes---i.e., the VEs are modelled as point transmitters,---while the dynamic DNT is modelled as discussed in Section \ref{sec:dynamic_dt_construction}. As expected, the localization performance increases with the level of supervision in both scenarios, with an enhancement of $\approx 11$ times between 5 \% and 50 \% supervision. 
The \gls{tsne} based method, tested on the realistic static and dynamic DNTs, achieves decreasing KS and overall high CT and TW over increasing supervision levels.

The more challenging features of the dynamic DNT with respect to the static one are highlighted by the reduction of both channel charting and localization performance for all the supervision levels. Whereas the charting performance is only slightly affected, the increases in terms of mean localization error range from $\approx 1.49$ m for 50 \% supervision up to $\approx 6.3$~m for 5 \% supervision between the static and dynamic cases.

\begin{figure}[!t]
    \centering
    \includegraphics[width=0.9\columnwidth]{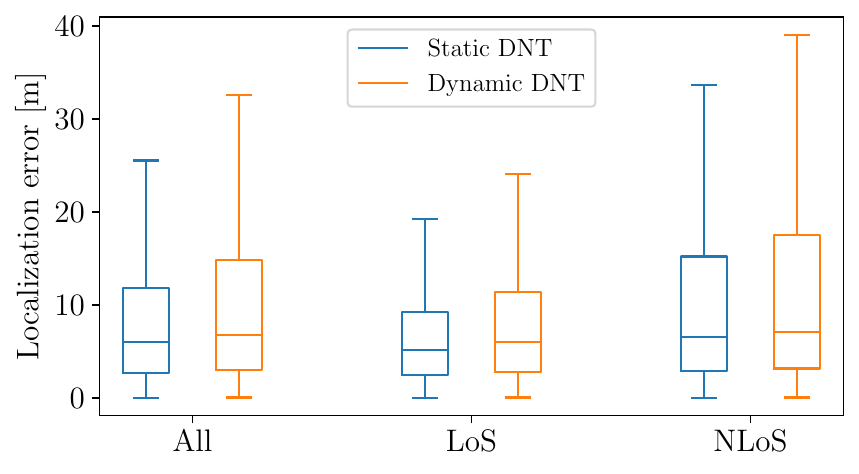}
    \vspace{-0.2cm}
    \caption{\textit{Chartwin} localization error box plots obtained for static and dynamic DNTs separating the data points in LoS and NLoS conditions. The proposed boxes present the localization performance on the validation set (composed by $\approx 40$\% LoS samples and $\approx60$\% NLoS samples) using labels semi-supervision for 25 \% of the data points.}
    \vspace{-0.2cm}
    \label{fig:loc_error_los_nlos_box_plots}
\end{figure}

These results are consistent with the empirical cumulative distribution function (ECDF) curves for the localization error proposed in Fig. \ref{fig:localization_error_ecdf}, where continuous lines represent the static DNT while dashed lines outline the results for the dynamic DNT. 
We compared the proposed \textit{Chartwin} with the RSSI fingerprinting-based localization method in \cite{AlkhateebLoc} and with the MUSIC-based method in \cite{mazokha2023mobloc}, trained over channels sampled on a grid with $4 \times 4$ m resolution and evaluated on points randomly sampled across the scenario. 

In Fig. \ref{fig:loc_error_los_nlos_box_plots}, we examine the localization error using box plots for $t$-SNE models trained with 25 \% supervision. In the figure, we separate the LoS and NLoS conditions of the VEs in the evaluation dataset. These more challenging features of dynamic radio propagation environments result in a localization error increase for both LoS and NLoS conditions. For the considered case at 25 \% supervision, in the static DNT the mean localization performance changes from $\approx 7$ m to $\approx 10.7$ m for LoS points and from $\approx 21.7$ m to $\approx 29.2$ m for NLoS points, with an almost doubled performance variation.

\section{Conclusion}\label{sec:conclusion}

In this work, we proposed \textit{Chartwin}, a case study on channel-based localization in V2X communications using semi-supervised channel charting over a dynamic DNT of an urban area of Milan, Italy integrating realistic vehicular traffic generation with accurate ray tracing channel simulation. We assess localization performance on both static and dynamic DNT conditions. Numerical results showcase the performance of semi-supervised channel charting in constructing a spatially consistent chart of the considered urban environment, fostering DNT-aided channel charting and localization.

\section*{Acknowledgment}
\footnotesize
The work was partially supported by the European Union under the Italian National Recovery and Resilience Plan (NRRP) of NextGenerationEU, partnership on “Telecommunications of the Future” (PE00000001 - program “RESTART”, Structural Project 6GWINET). This paper was also supported by the “Sustainable Mobility Center (Centro Nazionale per la Mobilità Sostenibile – CNMS)” project funded by the European Union NextGenerationEU program within the PNRR, Mission 4 Component 2 Investment 1.4. The authors acknowledge the support of the Laboratorio di Simulazione Urbana ``Fausto Curti" of the Department of Architecture and Urban Studies of Politecnico di Milano in the provision of the highly detailed urban digital twin maps of the area of Milan, Italy used in this work.

\bibliographystyle{IEEEtran}
\bibliography{Bibliography}

\end{document}